\documentclass[sigconf]{acmart}

\AtBeginDocument{
  }

\usepackage{makecell}
\usepackage{enumitem}

\setcopyright{acmlicensed}
\copyrightyear{2027}
\acmYear{2027}

\acmConference[WSDM '27]{The 20th ACM International Conference on Web Search and Data Mining (WSDM 2027)}{February 15--19, 2027}{Hong Kong, China}

\begin{document}

\title{BRIDGE: Bridging the Semantic Quantization Gap for Generative Recommendation}


\settopmatter{authorsperrow=1}

\author{\mbox{Yatong Xiao, Mingjie Sun, Peng Wang, Wenjin Wu, Peng Jiang}}
\authornote{Corresponding author.}

\affiliation{%
  \institution{Kuaishou Technology}
  \city{Beijing}
  \country{China}
}

\email{{xiaoyatong, sunmingjie, wangpeng16, wuwenjin, jiangpeng}@kuaishou.com}

\begin{abstract}

Generative recommendation retrieves items by autoregressively generating \emph{Semantic IDs} (SIDs), where an RQ-VAE quantizes each item into a hierarchical SID sequence that the decoder generates level by level. We identify an overlooked bottleneck in this paradigm: the \emph{Semantic Quantization Gap} (SQG). Quantization derives SIDs from rich continuous signals, including codebook centroids, local neighborhood geometry, and level-wise entropy. However, the decoder receives only discrete identifiers and loses the semantic structure that shaped them. This mismatch manifests in three ways: a \emph{grounding gap}, where the decoder lacks semantic anchors when target-side information is limited; a \emph{neighborhood gap}, where token-level supervision discards inter-item similarity and exacerbates exposure bias; and a \emph{difficulty gap}, where computation is allocated uniformly despite generation difficulty diminishing with depth.

To bridge the SQG, we propose \textbf{BRIDGE}. \emph{Codebook-grounded latent reasoning} injects frozen quantizer centroids as semantic anchors before SID generation. \emph{Privileged EMA self-distillation} transfers neighborhood-aware knowledge from a teacher to an on-policy student, preserving inter-item geometry while mitigating exposure bias. \emph{Effort-aware recurrence} allocates more computation to the challenging early levels and progressively reduces it with depth. Across three Amazon benchmarks, BRIDGE improves Recall@10 and NDCG@10 by 10.9\% and 9.7\% on average over the strongest baseline (up to 12.6\% and 10.7\%). In a one-week online A/B experiment on a large-scale e-commerce advertising platform with 5\% traffic, BRIDGE improves eCPM by 0.65\%.

\end{abstract}

\maketitle

\section{Introduction}
\label{sec:intro}
Generative recommendation has emerged as a promising paradigm that casts item retrieval as autoregressive sequence generation~\cite{rajput2023tiger,zheng2024adapting}. An RQ-VAE quantizer first maps each item into a hierarchical sequence of Semantic IDs (SIDs), where each level refines the representation of the previous one. At inference time, a decoder autoregressively generates SID tokens level by level, effectively searching the item catalog through structured generation. Compared with dual-tower retrieval, this paradigm explicitly models the semantic hierarchy of items and has demonstrated strong performance on standard benchmarks.

\begin{figure*}[t]
  \centering
  \includegraphics[width=\linewidth]{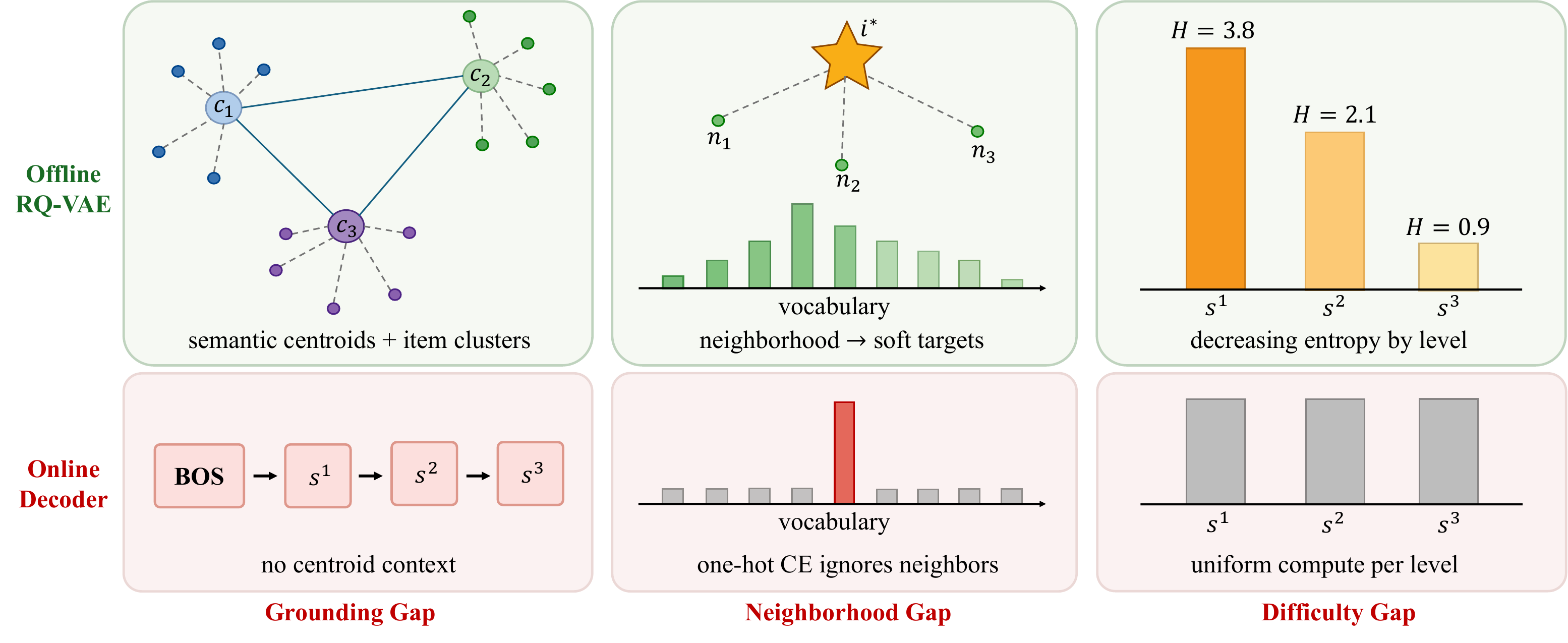}
  \caption{The Semantic Quantization Gap (SQG).}

  \label{fig:sqg}
\end{figure*}

Despite this progress, we identify a fundamental asymmetry between the quantization stage and the generation stage that has been largely overlooked. Quantization constructs SIDs from the rich continuous structure of item embeddings: codebook centroids encode coarse semantic clusters, inter-item neighborhood geometry captures relational proximity, and level-wise entropy reflects how ambiguous each level of the hierarchy is. None of this structure is accessible to the decoder, which receives only discrete code indices as supervision. We term this asymmetry the \emph{Semantic Quantization Gap} (SQG), illustrated in Figure~\ref{fig:sqg}, and observe that it manifests at three distinct points.

\textbf{Grounding gap.}
The first-level SID token $s^1$ partitions items into coarse semantic clusters and is the highest-entropy, hardest-to-predict decision—errors here cascade through all subsequent levels. Yet the decoder must predict $s^1$ with the least target-side context, having no access to the codebook centroids that defined those clusters in the first place. This makes the model prone to unstable coarse routing, especially when user histories are short or ambiguous.

\textbf{Neighborhood gap.}
One-hot supervision treats every incorrect SID as equally wrong, discarding the graded semantic proximity encoded in the quantizer's neighborhood geometry. Furthermore, because the decoder is trained on ground-truth SID prefixes but must generate its own at inference time, standard teacher forcing induces exposure bias—a train--test mismatch that one-hot targets cannot mitigate.

\textbf{Difficulty gap.}
Residual quantization by design makes early levels high-entropy and later levels increasingly deterministic refinements. A standard decoder allocates uniform compute across all levels, inverting the natural difficulty prior and wasting capacity on easy levels while underserving hard ones, which ultimately hurts top-rank accuracy where early-level mistakes are most costly.

We propose \textbf{BRIDGE}, a difficulty-aware framework that reinserts the discarded quantization structure at all three points. \emph{Codebook-grounded latent reasoning} (CGLR) injects frozen quantizer centroids as semantic anchors into a latent representation formed before $s^1$ generation, closing the grounding gap. \emph{Privileged EMA self-distillation} (PESD) trains an EMA teacher on semantically neighboring items and distills its soft distributions into the on-policy student, recovering neighborhood geometry and mitigating exposure bias. \emph{Effort-aware recurrence} (EAR) allocates more iterative refinement to early, high-entropy levels and tapers compute with depth, matching the difficulty prior. The three components couple architecturally: latent reasoning constitutes the first recurrent iteration, and its codebook-anchored representation serves simultaneously as the self-distillation teacher, avoiding redundant computation. In summary, our contributions are as follows:

\begin{itemize}[leftmargin=*,nosep]
  \item We identify and formalize the Semantic Quantization Gap (SQG), a previously overlooked encoder--decoder asymmetry in generative recommendation, and decompose it into three actionable gaps.
  \item We propose BRIDGE, which closes each gap with a targeted, architecturally coupled component: codebook-grounded latent reasoning (CGLR), privileged EMA self-distillation (PESD), and effort-aware recurrence (EAR).
  \item Experiments on three Amazon benchmarks (Sports, Beauty, and Toys) show that BRIDGE improves Recall@10 and NDCG@10 by 10.9\% and 9.7\% on average over strong generative retrieval baselines (up to 12.6\% and 10.7\%), with ablations indicating the independent contribution of each component.
\end{itemize}

\section{Related Works}

\subsection{Semantic ID Tokenization and Generative Recommendation}
Traditional sequential recommenders predict over arbitrary item IDs, which weakens semantic generalization. \citet{rajput2023tiger} proposes \textit{TIGER}, tokenizing each item into a short tuple of RQ-VAE Semantic IDs (SIDs) and training an encoder--decoder generator to autoregressively predict the next-item SID given user history. Subsequent work mainly improves tokenization quality: \textit{CoST} adds contrastive quantization to reduce collisions~\citep{zhu2024cost}; \textit{LETTER} learns codebooks jointly from semantics and CF signals~\citep{wang2024learnable}; \textit{TokenRec} studies SID tokenization tailored to LLM-style decoding~\citep{qu2025tokenrec}; and \textit{QARM} extends SID generation to multimodal inputs~\citep{luo2025qarm}. Industrial studies \textit{PLUM} and \textit{FORGE} further analyze large-scale deployment and evaluation of SID-based retrieval~\citep{he2026plum,fu2026forge}. Beyond content-only identifiers, \textit{ColaRec} and \textit{SEATER} inject collaborative structure via hybrid generation and tree-structured SIDs~\citep{wang2024content,si2024generative}. On the generator side, decoder-only Transformers are widely used after~\citet{rajput2023tiger}; \textit{LC-Rec} adapts LLMs by bridging the LLM vocabulary and SID space with a learnable codebook~\citep{zheng2024adapting}.

\subsection{Latent Reasoning in Generative Models}
Implicit/latent reasoning aims to retain multi-step computation while avoiding explicit chain-of-thought token overhead. Representative approaches include internal rationale tokens (\textit{Quiet-STaR}~\citep{zelikman2024quiet}), continuous hidden-state feedback (\textit{COCONUT}~\citep{hao2024training}), and distilling explicit CoT into weights (\textit{Implicit CoT}~\citep{deng2023implicit,deng2024explicit}). \citet{chen2026makes} analyzes latent-CoT failure modes (e.g., semantic drift) and proposes separating trajectory-versus-space supervision. In recommendation, latent intent reasoning has been explored via factorized continuous spaces (\textit{FLR}~\citep{gao2026factorized}) and hierarchy-scaffolded intent signals (\textit{CaLIR}~\citep{zhang2026beyond}). Different from these, CGLR explicitly anchors latent reasoning to the tokenizer geometry by projecting $\mathbf{r}$ into the RQ-VAE codebook space and regularizing it toward the ground-truth centroid.

\subsection{Knowledge Distillation with Privileged Information}
Learning Using Privileged Information (LUPI) formalizes the use of additional training-only signals through a teacher~\citep{vapnik2009new}. In recommendation, privileged information has been instantiated as richer graph context or side-information that improves student generalization when such context is unavailable at inference~\citep{wang2021privileged}. Self-distillation can also leverage training-only traces (e.g., verification signals) to refine a student~\citep{zhao2026self}. EMA teachers are widely used to stabilize self-training and avoid representational collapse (e.g., Mean Teacher~\citep{tarvainen2017mean} and DINO~\citep{caron2021emerging}). PESD combines these ideas for generative recommendation: the EMA teacher receives privileged semantic-neighbor context to produce geometry-aware soft targets, which are distilled to an on-policy student to reduce exposure bias.

\subsection{Adaptive and Recurrent Computation in Decoder-Only Models}
Adaptive-depth computation reallocates compute based on difficulty. Classic examples include ACT~\citep{graves2016adaptive}, and weight-tied recurrence in Universal Transformers~\citep{dehghani2018universal}. Looped Transformers further study fixed-weight iterative computation and inference-time scaling via repeated application of the same block~\citep{giannou2023looped,yang2024looped}. In contrast to uniform compute across positions, EAR uses a simple hierarchy-aware schedule ($m_1 \ge \cdots \ge m_L$) so that coarse, high-entropy SID levels receive more recurrent refinement while later residual levels are computed more cheaply.

\section{Method}

\subsection{Problem Formulation}

\paragraph{Sequential Recommendation}
Let $\mathcal{U}$ and $\mathcal{I}$ denote the user and item sets, respectively. For a target user $u$, we denote their interaction history as an ordered sequence $H_u = (i_1, i_2, \ldots, i_T)$, where $i_t \in \mathcal{I}$ is the item interacted with at time step $t$. The goal of sequential recommendation is to predict the next item $i^+ \in \mathcal{I}$ that $u$ is most likely to interact with.

\paragraph{Item Representation via RQ-VAE}
Each item $i$ is first encoded into a continuous semantic embedding $\mathbf{e}_i \in \mathbb{R}^d$ by a pretrained text, multimodal, or collaborative filtering encoder. To enable discrete generative modeling, we adopt Residual Quantization to convert $\mathbf{e}_i$ into an $L$-level \emph{Semantic ID} (SID):
\begin{equation}
  \mathbf{s}_i = \bigl(s_i^1,\, s_i^2,\, \ldots,\, s_i^L\bigr),
  \label{eq:sid}
\end{equation}
where $s_i^\ell \in \{1, \ldots, K\}$ is the discrete codebook index at level $\ell$. The residual quantization procedure is defined recursively as
\begin{align}
  \mathbf{r}_i^1 &= \mathbf{e}_i, \label{eq:rq_init}\\
  s_i^\ell &= \arg\min_{k}\,\bigl\|\mathbf{r}_i^\ell - \mathbf{c}_k^\ell\bigr\|_2, \label{eq:rq_assign}\\
  \mathbf{r}_i^{\ell+1} &= \mathbf{r}_i^\ell - \mathbf{c}_{s_i^\ell}^\ell, \label{eq:rq_residual}
\end{align}
where $\mathbf{c}_k^\ell$ is the $k$-th codeword in the $\ell$-th codebook $\mathcal{C}^\ell$. Each level quantizes the residual error of the previous level, so the concatenation $\mathbf{s}_i$ progressively refines the approximation of $\mathbf{e}_i$.

\paragraph{Autoregressive Generation Objective}
Given the user history $H_u$, the model predicts the SID of the target item $i^+$ token by token in an autoregressive manner:
\begin{equation}
  p\!\left(\mathbf{s}_{i^+} \mid H_u\right)
  = \prod_{\ell=1}^{L}
    p\!\left(s_{i^+}^\ell \;\middle|\; H_u,\, s_{i^+}^{<\ell}\right),
  \label{eq:ar}
\end{equation}
where $s_{i^+}^{<\ell} = (s_{i^+}^1, \ldots, s_{i^+}^{\ell-1})$ denotes all previously generated tokens.

\subsection{Overview of \textsc{Bridge}}

\label{sec:overview}

\begin{figure*}[t]
  \centering
  \includegraphics[width=\linewidth]{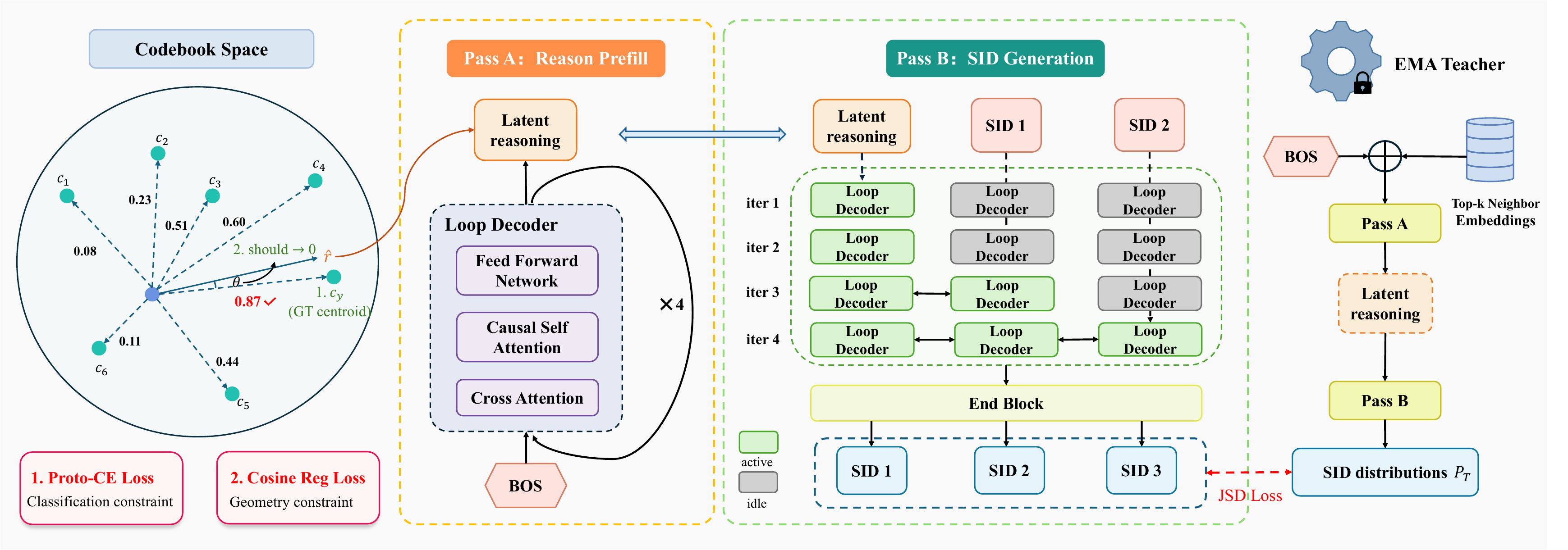}
  \caption{Overview of \textsc{Bridge}. User history embeddings are encoded into a context memory. A recurrent loop decoder then runs a hierarchical schedule: Pass~A iterates $R$ times over a reasoning slot to produce a latent representation $\mathbf{r}$; Pass~B generates SID tokens level by level with a right-aligned loop count, so coarser levels receive more iterations. At training time, an EMA teacher with privileged semantic neighbors provides soft targets via JSD distillation. Frozen codebook centroids anchor $\mathbf{r}$ throughout training.}
  \label{fig:overview}
\end{figure*}

Figure~\ref{fig:overview} illustrates the overall architecture of \textsc{Bridge}. We build on a generative retrieval backbone that encodes user history into a memory sequence and decodes a hierarchical SID via a shared recurrent loop block. \textsc{Bridge} augments this backbone with three coupled mechanisms that close the three gaps identified in Section~\ref{sec:intro}.

\paragraph{Backbone}
Given a user's interaction history embeddings $\{\mathbf{e}_{i_1}, \ldots, \mathbf{e}_{i_T}\}$, a two-layer MLP projects each embedding into model space and adds a learned positional embedding, yielding memory $\mathbf{M}_u \in \mathbb{R}^{T \times d}$. Both the reasoning pass and the SID generation pass share a single \emph{loop decoder} $\mathrm{LoopDec}$ and a joint input projection $\mathrm{InputProj}: \mathbb{R}^{2d} \to \mathbb{R}^d$. A positional slot embedding table $\mathbf{P} \in \mathbb{R}^{(L+1) \times d}$ assigns slot 0 to the reasoning position and slots $1, \ldots, L$ to SID levels.

\paragraph{Pass A — Reasoning (CGLR)}
Before any SID token is generated, \textsc{Bridge} runs a reasoning pass to produce a continuous latent representation $\mathbf{r} \in \mathbb{R}^d$. The slot-0 positional embedding and the learnable \texttt{[BOS]} embedding are concatenated and projected to initialize a single reasoning token: $\mathbf{h}^{(0)} = \mathrm{InputProj}([\mathbf{p}_0 \,\|\, \mathbf{b}])$. This token is refined by $R$ iterations of $\mathrm{LoopDec}$, with a dedicated per-iteration embedding $\delta^{(i)}_A$ from a learnable table $\mathbf{E}_A \in \mathbb{R}^{R \times d}$ injected at each step:
\begin{equation}
  \mathbf{h}^{(i)} = \mathrm{LoopDec}\!\left(
    \mathbf{h}^{(i-1)} + \boldsymbol{\delta}^{(i)}_A,\; \mathbf{M}_u
  \right), \quad i = 1, \ldots, R.
\end{equation}
The final state $\mathbf{r} = \mathbf{h}^{(R)}$ is anchored to frozen quantizer centroids via a dual codebook loss (Section~\ref{sec:cglr}), closing the \emph{grounding gap}.

\paragraph{Pass B — SID Generation (EAR)}
The reasoning representation $\mathbf{r}$ replaces \texttt{[BOS]} as the first input to the SID sequence. Each position $\ell \in \{1, \ldots, L\}$ is initialized by projecting its slot embedding concatenated with its content token:
\begin{equation}
  \mathbf{h}^{(0)}_\ell = \mathrm{InputProj}\!\left(
    [\mathbf{p}_\ell \,\|\, x_\ell]\right),
\end{equation}
where $x_1 = \mathbf{r}$. During training we use teacher forcing so $x_\ell = \mathrm{emb}(s^{\ell-1}_{i^+})$ for $\ell > 1$, while at inference time $x_\ell$ is the embedding of the previously \emph{generated} token. A right-aligned recurrent schedule with per-level budgets $\mathbf{m} = (m_1, \ldots, m_L)$, $m_1 \ge \cdots \ge m_L$, runs $\mathrm{LoopDec}$ over an expanding active prefix, allocating more iterations to harder early levels (Section~\ref{sec:loop}). A lightweight end block applied once after all iterations normalizes the level-specific hidden states before logit projection, closing the \emph{difficulty gap}.

\paragraph{Self-Distillation (PESD)}
An EMA shadow of the student serves as a privileged teacher: during training, the mean embedding of the target item's top-$k$ semantic neighbors is projected onto the teacher's \texttt{[BOS]}, giving the teacher access to neighborhood geometry invisible to the student. The student minimizes a symmetrized JSD between its on-policy logits and the teacher's soft distributions, recovering the quantizer's neighborhood structure and mitigating teacher forcing exposure bias (Section~\ref{sec:pesd}), closing the \emph{neighborhood gap}.

\paragraph{Coupling}
The three components share $\mathrm{LoopDec}$ and $\mathrm{InputProj}$, and interact through $\mathbf{r}$: the codebook-anchored reasoning representation from Pass A is reused as the distillation anchor in PESD, coupling grounding and neighborhood reinjection without additional computation. The overall training objective is summarized in Eq.~\eqref{eq:total_loss}.

\subsection{Codebook-Grounded Latent Reasoning (CGLR)}
\label{sec:cglr}

To bridge the semantic grounding gap, we propose \emph{Codebook-Grounded Latent Reasoning} (CGLR): before emitting any SID token, the shared recurrent loop decoder synthesizes a continuous latent reasoning representation that is explicitly anchored to frozen semantic codebook centroids. This turns the quantizer codebook—originally an offline tokenization artifact—into a generation-time semantic interface.

\subsubsection{Latent Reasoning Prefill}

Standard SID-based decoders condition generation directly on a \texttt{[BOS]} token, bypassing the continuous semantic structure of the item space. We prepend a dedicated reasoning pass (Pass~A) that runs \emph{before} any SID token is predicted.

\paragraph{Pass~A: Reasoning}
A learnable \texttt{[BOS]} embedding $\mathbf{b} \in \mathbb{R}^d$ is combined with a positional slot embedding $\mathbf{p}_0 \in \mathbb{R}^d$ and fused by a linear projection:
\begin{equation}
  \mathbf{h}^{(0)} = W_{\mathrm{in}}\bigl[\mathbf{p}_0 \| \mathbf{b}\bigr]
  \in \mathbb{R}^d,
  \label{eq:pass_a_init}
\end{equation}
where $\|\cdot\|$ denotes concatenation. The hidden state is then refined by $R$ iterations of the shared loop block, with a learnable iteration embedding $\boldsymbol{\delta}^{(i)}_{\mathrm{A}}$ injected at each step to distinguish reasoning iterations from SID-generation iterations:
\begin{equation}
  \mathbf{h}^{(i)} = \mathrm{LoopDec}\!\left(
    \mathbf{h}^{(i-1)} + \boldsymbol{\delta}^{(i)}_{\mathrm{A}},\;
    \mathbf{M}_u
  \right), \quad i = 1, \ldots, R.
  \label{eq:pass_a_loop}
\end{equation}
The final hidden state is taken as the latent reasoning representation:
\begin{equation}
  \mathbf{r} = \mathbf{h}^{(R)} \in \mathbb{R}^d.
  \label{eq:reason_repr}
\end{equation}
$\mathrm{LoopDec}$ cross-attends to the user context memory $\mathbf{M}_u$ at every iteration; no causal self-attention mask is needed since the reasoning slot is a single token. By design $R \ge \max_\ell(\text{loop\_count}_\ell)$, ensuring the reasoning slot receives at least as many iterations as any SID level.

\paragraph{Pass~B: SID Generation}
The reasoning representation $\mathbf{r}$ replaces \texttt{[BOS]} as the first input to the SID decoder (detailed in Section~\ref{sec:loop}), so that every SID token is conditioned on the refined semantic reasoning:
\begin{equation}
  \mathrm{input} = \bigl[\mathbf{r},\;
    \mathrm{emb}(s^1_{i^+}),\; \ldots,\; \mathrm{emb}(s^{L-1}_{i^+})\bigr].
  \label{eq:pass_b_input}
\end{equation}
The representation $\mathbf{r}$ is \emph{not detached}: gradients from both the SID generation loss and the codebook anchoring objective propagate back through Pass~A, jointly shaping the reasoning representation.

\subsubsection{Codebook Anchoring Objective}

Without explicit constraints, $\mathbf{r}$ can drift arbitrarily in latent space, decoupling from the geometric structure of the item embedding space. We anchor $\mathbf{r}$ to a frozen semantic codebook via a coarse semantic label that operates at a \emph{broader granularity} than the RQ-VAE quantization codes. The coarse label is used only for training-time supervision and is never provided to the model at inference time.

\paragraph{Coarse Semantic Label}
Rather than using the first-level RQ-VAE code $s^1_{i^+}$ as the anchoring target, we derive a coarser cluster index $z \in \{0, \ldots, V_c{-}1\}$ by \emph{re-clustering} the \emph{level-1 codebook centroids} (not item embeddings), where $V_c \ll K$. Concretely, we run an offline clustering procedure (e.g., $k$-means) over $\{\mathbf{v}^{(1)}_k\}_{k=1}^K$ and obtain coarse prototypes $\boldsymbol{\mu}_0,\ldots,\boldsymbol{\mu}_{V_c-1} \in \mathbb{R}^d$. Each item is then assigned to the coarse cluster of its level-1 centroid:
\begin{equation}
  z_i = \arg\min_{k} \left\|\mathbf{v}^{(1)}_{s^1_i} - \boldsymbol{\mu}_k\right\|_2,
  \label{eq:coarse_cluster}
\end{equation}
where $k \in \{0,\ldots,V_c-1\}$ and $\mathbf{v}^{(1)}_{s^1_i}$ is the level-1 centroid indexed by $s^1_i$. This label groups items at a broader semantic granularity than the RQ-VAE quantization hierarchy, providing a more stable anchor: it encodes \emph{which broad semantic category} the target item belongs to rather than its precise quantization code. The corresponding frozen coarse-prototype set $\mathcal{M} = \{\boldsymbol{\mu}_0, \ldots, \boldsymbol{\mu}_{V_c-1}\}$ is precomputed once and kept fixed during recommendation training.

\paragraph{Anchoring Losses}
A lightweight projection $g(\cdot)$ maps $\mathbf{r}$ into the prototype space, followed by $\ell_2$-normalization:
\begin{equation}
  \mathbf{q} = \frac{g(\mathbf{r})}{\|g(\mathbf{r})\|_2}, \qquad
  \hat{\boldsymbol{\mu}}_k = \frac{\boldsymbol{\mu}_k}{\|\boldsymbol{\mu}_k\|_2}.
  \label{eq:reason_proj}
\end{equation}
We apply two complementary losses.

\emph{Prototype classification} enforces global discriminative separation across all coarse prototypes:
\begin{equation}
  \mathcal{L}_{\mathrm{proto}} =
    -\log \frac{\exp\!\left(\mathbf{q}^\top \hat{\boldsymbol{\mu}}_z / \tau\right)}
               {\displaystyle\sum_{k=0}^{V_c-1}
                \exp\!\left(\mathbf{q}^\top \hat{\boldsymbol{\mu}}_k / \tau\right)},
  \label{eq:proto}
\end{equation}
where $\tau$ is a temperature hyperparameter.

\emph{Cosine regression} provides a dense directional pull toward the target prototype, complementing the classification boundaries.
\begin{equation}
  \mathcal{L}_{\mathrm{reg}} = 1 - \mathbf{q}^\top \hat{\boldsymbol{\mu}}_z.
  \label{eq:reg}
\end{equation}

The two losses work well together: prototype CE helps separate clusters but does not guarantee that $\mathbf{r}$ points toward the right prototype; cosine regression pulls $\mathbf{r}$ toward the target direction but is weaker at enforcing separation. Using both keeps $\mathbf{r}$ close to its prototype and away from others.

We combine them as the CGLR loss:
\begin{equation}
  \mathcal{L}_{\mathrm{CGLR}} = \lambda_{\mathrm{proto}}\,\mathcal{L}_{\mathrm{proto}} + \lambda_{\mathrm{reg}}\,\mathcal{L}_{\mathrm{reg}}.
  \label{eq:lcglr}
\end{equation}

The prototype set $\mathcal{M}$ is frozen throughout training; no additional item-side annotation is required.

\subsubsection{Codebook as a Training-Time Semantic Interface}

The frozen prototypes $\mathcal{M}$ encode the coarse semantic geometry of the item space: each prototype $\boldsymbol{\mu}_k$ represents a broad semantic cluster above the fine-grained RQ-VAE quantization.

\textsc{Bridge} uses this structure to define \emph{training-time} anchoring objectives: Eqs.~\eqref{eq:proto}--\eqref{eq:reg} encourage the reasoning representation $\mathbf{r}$ to encode \emph{which broad semantic category the target item belongs to} before the first SID token is predicted. At inference time, the model uses neither the coarse label $z$ nor any target-side prototype; it only uses the predicted $\mathbf{r}$ produced from user history.

Crucially, operating at coarse granularity makes the anchor more robust than supervising with the exact RQ-VAE code: coarse cluster membership is stable across minor embedding perturbations, whereas fine-grained quantization codes can flip between neighbors, producing an inconsistent training signal. This mechanism requires no changes to the tokenizer and applies to any generative recommender built on residual-quantization SIDs.

\subsection{Privileged EMA Self-Distillation (PESD)}
\label{sec:pesd}

To mitigate the \emph{neighborhood gap}, we introduce \emph{Privileged EMA Self-Distillation} (PESD). One-hot supervision discards graded semantic proximity among SIDs, and teacher forcing further amplifies the train--inference mismatch. PESD equips an EMA teacher with privileged semantic neighbors to produce semantically-aware soft targets, which are distilled into an on-policy student to recover neighborhood geometry and improve robustness under self-generated prefixes.

\subsubsection{Privileged Teacher and On-Policy Student}

\paragraph{EMA Teacher}
The teacher is an exponential moving average (EMA) shadow of the student, sharing the same architecture and weights but with $\texttt{requires\_grad}=\texttt{False}$. Its parameters are never updated by backpropagation; they evolve solely through the EMA update
  \begin{equation}
    \boldsymbol{\theta}_t \leftarrow m\,\boldsymbol{\theta}_t
      + (1-m)\,\boldsymbol{\theta}_s,
    \label{eq:ema}
  \end{equation}
where $m$ is the momentum coefficient and $\boldsymbol{\theta}_s$ are the student weights. The slowly-moving teacher provides stable soft targets and avoids the collapse that occurs when the student distills from itself directly~\cite{caron2021emerging}.

\paragraph{On-Policy Student}
For the distillation path, the student generates SID logits \emph{on policy}: at each level $\ell$ it takes the $\arg\max$ of its own previous logit as the next input, without teacher forcing. This closes the train--inference distribution gap and ensures that the distillation signal corrects errors under the same conditions the model faces at inference time. The teacher, by contrast, always uses teacher forcing inputs, so its logits reflect the ideal conditional distribution given correct prefixes.

\subsubsection{Semantic Neighbor Privilege}

\paragraph{Privilege Construction}
For each training example, we retrieve the top-$k$ semantic neighbors of the target item (excluding the target itself) in the original item-embedding space, and inject their mean embedding into the teacher's reasoning pass:
\begin{align}
  \bar{\mathbf{e}}_{\mathrm{nbr}} &= \frac{1}{k}\sum_{j=1}^{k}\mathbf{e}_{n_j},
  \label{eq:nbr_mean}\\
  \mathbf{o} &= \frac{W_{\mathrm{priv}}\,\bar{\mathbf{e}}_{\mathrm{nbr}}}
    {\|W_{\mathrm{priv}}\,\bar{\mathbf{e}}_{\mathrm{nbr}}\|_2}
    \cdot \gamma,
  \label{eq:priv_proj}
\end{align}
We compute neighbors using only the frozen item embeddings (no interaction labels), and we build the neighbor index on the training-time item catalog to avoid any split-dependent leakage. Here $W_{\mathrm{priv}}$ is a learnable projection and $\gamma$ is a learnable scale. The teacher's \texttt{[BOS]} embedding is then offset by $\mathbf{o}$ before Pass~A, so the teacher's reason representation $\mathbf{r}_t$ is conditioned on the semantic neighborhood of the target item.

\paragraph{Why Neighbors, Not Ground-Truth SID}
Providing the ground-truth SID directly to the teacher would collapse its output into a near-one-hot distribution, reducing distillation to a noisy copy of the hard CE loss. Neighbor embeddings encode the \emph{metric structure} of the item space—which items are semantically similar and to what degree—without revealing the exact target identity. The teacher's soft logits therefore carry information about the local semantic neighborhood, not just the single correct answer.

\paragraph{Privileged Encoder Training}
$W_{\mathrm{priv}}$ and $\gamma$ are trained exclusively by a teacher CE loss:
\begin{equation}
  \mathcal{L}_{\mathrm{tch}} =
    \frac{1}{L}\sum_{\ell=1}^{L}
    \mathrm{CE}\!\left(p_t^\ell,\, s^\ell_{i^+}\right),
  \label{eq:teacher_ce}
\end{equation}
which asks the privileged signal to be useful for predicting the target, rather than merely easy for the student to imitate. Critically, $W_{\mathrm{priv}}$ and $\gamma$ live on the student side; the teacher consumes their output but does not own them. This keeps the EMA stop-gradient property intact while still allowing the privileged encoder to learn through the teacher's activation graph. $\mathcal{L}_{\mathrm{tch}}$ is added to the total loss \emph{outside} the distillation weight ramp so the privileged signal is useful from the start.

\subsubsection{Distillation Losses and Stabilization}

\paragraph{SID Distillation}
Let $p_t^\ell$ and $p_s^\ell$ denote the teacher and on-policy student distributions at SID level $\ell$, obtained by applying softmax with temperature $T$ to the respective logits. We minimize their symmetrized Jensen--Shannon divergence:
\begin{equation}
  \mathcal{L}_{\mathrm{sid\text{-}dist}} =
    \frac{1}{L}\sum_{\ell=1}^{L}
    \widetilde{\mathrm{JSD}}_\alpha\!\left(p_t^\ell,\, p_s^\ell\right),
  \label{eq:jsd}
\end{equation}
\begin{equation}
  \widetilde{\mathrm{JSD}}_\alpha(p, q) =
    \frac{\alpha\,\mathrm{KL}(p \| m) + (1-\alpha)\,\mathrm{KL}(q \| m)}
         {\log V},
  \quad m = \alpha p + (1-\alpha)q,
  \label{eq:jsd_def}
\end{equation}
where $V$ is the vocabulary size and the $\log V$ normalization keeps the loss in $[0, 1]$ regardless of vocabulary size. Teacher logits are detached in Eq.~\eqref{eq:jsd}: minimizing $\mathrm{JSD}(\text{student} \| \text{teacher})$ with a live teacher graph admits a degenerate solution in which the teacher collapses toward the student, destroying the privileged signal.

\paragraph{Reason Distillation}
To align the student's latent reasoning with the privileged teacher's, we add a cosine distillation term on the reason representations:
\begin{equation}
  \mathcal{L}_{\mathrm{rsn\text{-}dist}} =
    1 - \frac{\mathbf{r}_s}{\|\mathbf{r}_s\|_2}
        \cdot \frac{\mathrm{sg}(\mathbf{r}_t)}{\|\mathbf{r}_t\|_2},
  \label{eq:reason_dist}
\end{equation}
where $\mathrm{sg}(\cdot)$ denotes stop-gradient. We reuse the $\mathbf{r}_s$ computed in the main forward pass rather than re-running Pass~A, so that the distillation gradient acts on the same representation that conditions SID generation.

\paragraph{Entropy Gate}
Early in training, the teacher's distributions can be over-confident or unreliable. We gate the distillation losses with:
\begin{equation}
  \mathcal{L}_{\mathrm{dist}} =
  \begin{cases}
    \lambda_r\,\mathcal{L}_{\mathrm{rsn\text{-}dist}}
    + \mathcal{L}_{\mathrm{sid\text{-}dist}} &
    \text{if } \min_\ell \mathcal{H}(p_t^\ell) > \epsilon, \\
    0 & \text{otherwise,}
  \end{cases}
  \label{eq:entropy_gate}
\end{equation}
where $\mathcal{H}$ is Shannon entropy and $\epsilon$ is a threshold hyperparameter. The gate is evaluated on detached logits and never contributes to any gradient graph.

We combine the PESD objectives as:
\begin{equation}
  \mathcal{L}_{\mathrm{PESD}} = \lambda_{\mathrm{tch}}\,\mathcal{L}_{\mathrm{tch}} + w(t)\,\mathcal{L}_{\mathrm{dist}}.
  \label{eq:lpesd}
\end{equation}

\paragraph{Distillation ramp}
The distillation weight follows a linear warm-up:
\begin{equation}
  w(t) = \min\!\left(1,\;\frac{t}{T_{\mathrm{warm}}}\right) \cdot w_{\max},
  \label{eq:ramp}
\end{equation}
where $t$ is the current training step and $T_{\mathrm{warm}}$ is the ramp duration. This delays distillation until the student's own representations are stable enough to benefit from the teacher signal.

\subsection{Effort-Aware Hierarchical Recurrence (EAR)}
\label{sec:loop}

Residual quantization induces a structured difficulty hierarchy across SID levels: the first level $s^1$ selects a coarse semantic region from a high-entropy distribution over the full item space, while later levels perform residual refinement conditioned on an increasingly constrained prefix, resulting in progressively lower prediction entropy. Standard autoregressive decoders apply uniform computation to every SID position, ignoring this prior. We address the hierarchical difficulty gap with \emph{effort-aware hierarchical recurrence}: a non-uniform recurrent refinement budget that allocates more iterations to early, high-entropy levels and fewer to later, easier ones.

\subsubsection{Right-Aligned Loop Schedule}
Let $\mathbf{m} = (m_1, m_2, \ldots, m_L)$ be the per-level loop counts, with the non-increasing constraint $m_1 \ge m_2 \ge \cdots \ge m_L > 0$. All levels share a single loop block $\mathrm{LoopDec}$ with the same weights. Pass~B runs for $M = m_1$ total iterations; at each iteration $j \in \{1,\ldots,M\}$ a \emph{right-aligned} schedule determines which prefix of positions is active:
\begin{equation}
  \mathcal{A}(j) = \bigl\{\ell \in \{1,\ldots,L\} : j \ge M - m_\ell + 1\bigr\}.
  \label{eq:active_set}
\end{equation}
Only positions in $\mathcal{A}(j)$ are updated at iteration $j$; positions outside it remain unchanged. The update for the active prefix is:
\begin{equation}
  \mathbf{h}^{(j)}_{\mathcal{A}(j)} = \mathrm{LoopDec}\!\left(
    \mathbf{h}^{(j-1)}_{\mathcal{A}(j)} + \boldsymbol{\delta}^{(j)}_{\mathrm{B}},\;
    \mathbf{M}_u;\;
    \mathrm{mask}_{\mathcal{A}(j)}
  \right),
  \label{eq:loop_update}
\end{equation}
where $\boldsymbol{\delta}^{(j)}_{\mathrm{B}}$ is a learnable iteration embedding, and $\mathrm{mask}_{\mathcal{A}(j)}$ is the causal mask restricted to the active prefix.

Right-alignment delays the participation of deeper SID levels so that when a new level $\ell$ first enters $\mathcal{A}(j)$, the earlier prefix $\{1,\ldots,\ell-1\}$ has already undergone more recurrent refinements. Through causal self-attention within $\mathrm{LoopDec}$, level $\ell$ can therefore immediately condition on these better-refined hidden states of preceding SID levels, enabling cleaner residual refinement than attending to unrefined initial projections.

\paragraph{Example}
For $L=3$ and $\mathbf{m}=(4,2,1)$, the active sets across $M=4$ iterations are:
\begin{center}
\begin{tabular}{ccc}
\toprule
Iteration $j$ & Active positions $\mathcal{A}(j)$ & Cumulative refinements \\
\midrule
1 & $\{1\}$     & level 1: 1 \\
2 & $\{1\}$     & level 1: 2 \\
3 & $\{1,2\}$   & level 1: 3,\ level 2: 1 \\
4 & $\{1,2,3\}$ & level 1: 4,\ level 2: 2,\ level 3: 1 \\
\bottomrule
\end{tabular}
\end{center}

Level~1 participates in all 4 iterations, level~2 in 2, and level~3 in only 1. Right-alignment ensures that when level~$\ell$ joins the loop at iteration $j$, it can immediately attend to the already-refined hidden states of all earlier levels via the causal mask, rather than unrefined initial projections.

\subsubsection{End Block}

After $M$ recurrent iterations, positions have experienced different numbers of refinements and may carry heterogeneous hidden distributions. A lightweight non-recurrent end block $\mathrm{EndDec}$ is applied once to the full sequence:
\begin{equation}
  \mathbf{h}^{\mathrm{out}} = \mathrm{EndDec}\!\left(
    \mathbf{h}^{(M)},\; \mathbf{M}_u;\; \mathrm{mask}_{1:L}
  \right).
  \label{eq:end_block}
\end{equation}
This serves two purposes: (i) it normalizes the level-specific hidden distributions into a consistent output space; (ii) later positions attend one final time to the maximally-refined earlier prefix before producing their logits. The end block adds a fixed, one-time cost independent of $M$.

\subsubsection{Complexity Analysis}

Let $\Phi$ denote the per-token per-layer FLOPs of $\mathrm{LoopDec}$ and $N$ its number of layers. The total Pass~B recurrent cost is proportional to
\begin{equation}
  \Phi \cdot N \cdot \sum_{\ell=1}^{L} m_\ell,
  \label{eq:flops_loop}
\end{equation}
since $\mathrm{LoopDec}$ is applied once per active token per iteration. A standard decoder of depth $D$ applied uniformly to all $L$ positions costs $\Phi \cdot D \cdot L$. Under the schedule $\mathbf{m}=(4,2,1)$ with $N=3$ loop layers ($\sum m_\ell=7$, $\Phi N \sum m_\ell = 21\Phi$) versus a 9-layer uniform decoder ($\Phi \cdot 9 \cdot 3 = 27\Phi$), the loop variant achieves a $\approx$22\% FLOPs reduction while reallocating compute toward harder positions.

Critically, because $\mathrm{LoopDec}$ is \emph{shared} across all levels and all iterations, increasing $m_1$ does not introduce additional parameters—only additional forward passes through the same weights. The total parameter count is determined by the loop block depth $N$ and the single end block, not by $\mathbf{m}$.
\paragraph{EAR loss}
EAR is trained with the standard next-token prediction objective, plus a light embedding regularization term that we apply to the backbone across all variants (i.e., it is not specific to the EAR recurrence mechanism):
\begin{equation}
  \mathcal{L}_{\mathrm{EAR}} = \mathcal{L}_{\mathrm{ntp}} + \lambda_{l_2}\,\mathcal{L}_{l_2},
  \label{eq:lear}
\end{equation}
where $\mathcal{L}_{\mathrm{ntp}}$ is the next-token cross-entropy objective in Eq.~\eqref{eq:ar}, and $\mathcal{L}_{l_2}=\tfrac12\mathbb{E}\|\mathbf{x}_{\mathrm{tgt}}\|_2^2$ regularizes the teacher-forcing SID input embeddings.

\subsection{Training Objective}

The full \textsc{Bridge} training objective is the sum of component-wise losses:
\begin{equation}
  \mathcal{L} = \mathcal{L}_{\mathrm{EAR}} + \mathcal{L}_{\mathrm{CGLR}} + \mathcal{L}_{\mathrm{PESD}},
  \label{eq:total_loss}
\end{equation}
where $\mathcal{L}_{\mathrm{EAR}}$, $\mathcal{L}_{\mathrm{CGLR}}$, and $\mathcal{L}_{\mathrm{PESD}}$ are defined in Eqs.~\eqref{eq:lear}, \eqref{eq:lcglr}, and \eqref{eq:lpesd}, respectively. Our ablations disable $\mathcal{L}_{\mathrm{CGLR}}$ and/or $\mathcal{L}_{\mathrm{PESD}}$ while keeping $\mathcal{L}_{\mathrm{EAR}}$ unchanged.

\section{Experiments}

\subsection{Experimental Setup}

\paragraph{Datasets}
We evaluate on three public benchmarks: Amazon Beauty, Amazon Sports, and Amazon Toys. All datasets apply 5-core filtering and leave-one-out splitting: the last interaction of each user is held out for testing, the second-to-last for validation, and the remainder for training. Dataset statistics are reported in Table~\ref{tab:datasets}.

\begin{table}[h]
  \centering
  \caption{Dataset statistics after 5-core filtering.}
  \label{tab:datasets}
  \small
  \setlength{\tabcolsep}{3pt}
  \resizebox{\columnwidth}{!}{
  \begin{tabular}{lccccc}
    \toprule
    \textbf{Dataset} & \textbf{\#Users} & \textbf{\#Items}
      & \textbf{\#Interactions} & \textbf{Avg.Len} & \textbf{Sparsity} \\
    \midrule
    Beauty  & 22,363 & 12,101 & 198,502 & 8.88  & 99.93\% \\
    Sports  & 35,598 & 18,357 & 296,175 & 8.32  & 99.95\% \\
    Toys    & 19,412 & 11,924 & 167,526 & 8.63  & 99.93\% \\
    \bottomrule
  \end{tabular}}
\end{table}

\paragraph{SID Construction}
Item text and metadata are encoded with Qwen2.5-1.5B to obtain 1536-dimensional embeddings. A three-level RQ-VAE ($L=3$, codebook size $K=2048$ per level) is then applied to produce SIDs. The coarse semantic centroids ($V_c=256$) used for codebook anchoring are obtained by re-clustering the level-1 RQ-VAE centroids (e.g., via $k$-means) and kept frozen throughout recommendation training.

\paragraph{Metrics}
We report Recall@$\{5,10\}$ and NDCG@$\{5,10\}$ as primary retrieval metrics. To diagnose generation behavior we additionally report SID hit rate (the fraction of examples where the model's beam-search output matches the ground-truth SID at each level).

\paragraph{Baselines}
We compare \textsc{Bridge} against baselines from two categories.

\textit{Item ID-based methods:}
\begin{itemize}[leftmargin=*]
  \item \textbf{GRU4Rec}~\cite{hidasi2015session}: A GRU-based sequential recommender that models user interaction sequences with recurrent networks.
  \item \textbf{SASRec}~\cite{kang2018self}: A self-attention sequential recommender that captures long-range item dependencies via causal transformer layers.
  \item \textbf{BERT4Rec}~\cite{sun2019bert4rec}: A bidirectional transformer sequential recommender trained with a masked item prediction objective.
\end{itemize}

\textit{Semantic ID-based methods:}
\begin{itemize}[leftmargin=*]
  \item \textbf{TIGER}~\cite{rajput2023tiger}: foundational SID-based generative retrieval framework that tokenizes items with an RQ-VAE codebook into hierarchical SIDs and trains an autoregressive decoder to generate SID tuples conditioned on user history.

  \item \textbf{ReaRec}~\cite{tang2026think}: SID-based sequential recommender that introduces implicit multi-step reasoning in the decoder by iteratively feeding hidden states back as inputs before SID generation.

  \item \textbf{COBRA}~\cite{yang2026sparse}: unified generative recommender that combines sparse semantic IDs with dense representations via a cascaded sparse--dense architecture, and improves decoding efficiency with beam fusion for top-$K$ generation.

  \item \textbf{GR4AD}~\cite{xue2026generative}: production generative recommender for large-scale advertising at Kuaishou, featuring UA-SID tokenization that encodes advertising business signals, a LazyAR decoder for efficient multi-candidate generation, and value-aware supervised learning aligned with business objectives.
\end{itemize}

\paragraph{Implementation Details}
All models use $d=512$, 8 attention heads, and FFN dimension 2048. User history is truncated to 32 items; beam width is 20. \textsc{Bridge} uses a 3-layer loop block, the EAR schedule $\mathbf{m}=(4,2,1)$, and $R=4$ reasoning iterations. The EMA decay is $\alpha=0.999$. Privileged neighbors use the top-$k=3$ items by embedding cosine similarity. Distillation hyperparameters: temperature $T=1.0$, JSD $\alpha=0.5$, entropy gate $\epsilon=0.5$, reason distillation weight $\lambda_r=0.5$, codebook loss weights $\lambda_{\mathrm{proto}}=\lambda_{\mathrm{reg}}=0.1$, $\lambda_{l_2}=10^{-3}$, privileged encoder weight $\lambda_{\mathrm{tch}}=0.5$. Distillation is applied after a 2{,}000-step warm-up and ramped linearly over 3{,}000 steps to its maximum weight $w_{\max}=1.0$. All models are trained with AdamW and a cosine learning rate schedule with linear warm-up.

\subsection{Main Results}

\begin{table*}[t]
  \centering
  \caption{Main results on three benchmarks. Best in \textbf{bold}, second-best \underline{underlined}.}
  \label{tab:main}
  \renewcommand{\arraystretch}{1.05}
  \resizebox{\linewidth}{!}{
  \begin{tabular}{l *{4}{c} *{4}{c} *{4}{c}}
    \toprule
    & \multicolumn{4}{c}{\textbf{Sports}}
    & \multicolumn{4}{c}{\textbf{Beauty}}
    & \multicolumn{4}{c}{\textbf{Toys}} \\
    \cmidrule(lr){2-5}\cmidrule(lr){6-9}\cmidrule(lr){10-13}
    \textbf{Methods}
    & \makecell{R@5} & \makecell{N@5} & \makecell{R@10} & \makecell{N@10}
    & \makecell{R@5} & \makecell{N@5} & \makecell{R@10} & \makecell{N@10}
    & \makecell{R@5} & \makecell{N@5} & \makecell{R@10} & \makecell{N@10} \\
    \midrule
    \multicolumn{13}{c}{\textit{Item ID-based}} \\
    \midrule
    GRU4Rec~\cite{hidasi2015session}  & 0.0129 & 0.0086 & 0.0204 & 0.0110 & 0.0164 & 0.0099 & 0.0283 & 0.0137 & 0.0097 & 0.0059 & 0.0176 & 0.0084 \\
    SASRec~\cite{kang2018self}   & 0.0233 & 0.0154 & 0.0350 & 0.0192 & 0.0387 & 0.0249 & 0.0605 & 0.0318 & 0.0463 & 0.0306 & 0.0675 & 0.0374 \\
    BERT4Rec~\cite{sun2019bert4rec} & 0.0115 & 0.0075 & 0.0191 & 0.0099 & 0.0203 & 0.0124 & 0.0347 & 0.0170 & 0.0116 & 0.0071 & 0.0203 & 0.0099 \\
    \midrule
    \multicolumn{13}{c}{\textit{Semantic ID-based}} \\
    \midrule
    TIGER~\cite{rajput2023tiger}      & 0.0264 & 0.0181 & 0.0400 & 0.0225 & 0.0454 & 0.0321 & 0.0648 & 0.0384 & 0.0521 & 0.0371 & 0.0712 & 0.0432 \\
    ReaRec~\cite{tang2026think}     & 0.0214 & 0.0116 & 0.0332 & 0.0154 & 0.0450 & 0.0262 & 0.0704 & 0.0344 & 0.0523 & 0.0298 & 0.0764 & 0.0376\\
    COBRA~\cite{yang2026sparse} & 0.0305 & 0.0215 & 0.0434 & 0.0257 & 0.0537 & 0.0395 & 0.0725 & 0.0456 & 0.0619 & 0.0462 & 0.0781 & 0.0515 \\
    GR4AD~\cite{xue2026generative}        & \underline{0.0318} & \underline{0.0221} & \underline{0.0443} & \underline{0.0263} & \underline{0.0545} & \underline{0.0407} & \underline{0.0742} & \underline{0.0470} & \underline{0.0624} & \underline{0.0463} & \underline{0.0788} & \underline{0.0518}  \\
    \midrule
    \textsc{Bridge} (ours)
      & \textbf{0.0340} & \textbf{0.0239} & \textbf{0.0478} & \textbf{0.0284}
      & \textbf{0.0607} & \textbf{0.0443} & \textbf{0.0835} & \textbf{0.0520}
      & \textbf{0.0684} & \textbf{0.0501} & \textbf{0.0886} & \textbf{0.0572} \\
    Improv.
      & +6.9\% & +8.2\% & +7.8\% & +8.0\%
      & +11.3\% & +8.8\% & +12.6\% & +10.7\%
      & +9.6\% & +8.1\% & +12.4\% & +10.5\% \\
    \bottomrule
  \end{tabular}}
\end{table*}

Table~\ref{tab:main} reports results on three Amazon benchmarks. \textsc{Bridge} achieves the best performance on all metrics across all datasets, showing consistent improvements over strong discriminative and generative baselines.

\paragraph{Semantic ID-based vs.\ item ID-based}
Semantic ID-based methods consistently outperform item ID-based baselines (GRU4Rec, SASRec, BERT4Rec), suggesting that autoregressive generation over hierarchical SIDs provides a stronger inductive bias for sequential recommendation than dot-product matching over item indices. The gap is especially pronounced on Recall@10, where the structured SID hierarchy allows generative models to explore a broader, semantically coherent candidate space compared to nearest-neighbor retrieval.

\paragraph{Bridge vs.\ generative baselines}
Among semantic ID-based generative methods, \textsc{Bridge} achieves the best result on every reported dataset and metric, outperforming the evaluated generative baselines, including GR4AD. Against the strongest generative baseline, \textsc{Bridge} achieves relative improvements of 7.8\% / 12.6\% / 12.4\% in Recall@10 and 8.0\% / 10.7\% / 10.5\% in NDCG@10 on Sports, Beauty, and Toys, respectively. TIGER decodes SIDs directly from a \texttt{[BOS]} token without any semantic grounding, leaving the quantizer's continuous structure inaccessible to the decoder; this is precisely the grounding gap that \textsc{Bridge}'s \emph{codebook-grounded latent reasoning} (CGLR) closes. GR4AD focuses on efficient multi-candidate generation and business-aligned training, but it does not explicitly exploit the residual-quantization-induced level-wise difficulty prior. \textsc{Bridge}'s \emph{effort-aware recurrence} (EAR), which allocates more refinement iterations to high-entropy early levels, provides a complementary inductive bias that standard level-uniform decoding does not capture.

\paragraph{Effect of dataset scale}
Gains are more pronounced on Beauty and Toys than on Sports. All three datasets share near-identical sparsity ($\approx$99.93--99.95\%), so we attribute the difference to catalog size: Sports contains approximately 50\% more items than Beauty or Toys, expanding the SID prediction space and making it harder for any single model to realize proportional gains. On smaller, more focused catalogs, codebook-anchored latent reasoning injects semantic structure that more directly covers the item space, while privileged EMA self-distillation recovers neighborhood geometry from item embeddings rather than observed interactions; this signal's coverage depends primarily on the catalog, not on the number of observed interactions. The relatively smaller gain on Sports suggests that at larger catalog scale, the marginal benefit of semantic anchoring diminishes as the codebook must partition a wider embedding manifold.

\paragraph{Consistency across metrics}
Gains are consistent across both Recall and NDCG at both cut-offs, indicating that \textsc{Bridge} improves both coverage (retrieving more relevant items within the top-$K$) and ranking quality (placing relevant items higher within the retrieved set). The improvement on NDCG is generally comparable to or larger than that on Recall, suggesting that the codebook-grounded reasoning and effort-aware recurrence together benefit precision at early ranks rather than merely expanding tail coverage. Detailed component analysis is presented in Section~\ref{sec:ablation}.

\subsection{Ablation Study}
\label{sec:ablation}

Table~\ref{tab:ablation} reports an ablation study on Beauty, comparing the EAR backbone with variants that add CGLR and/or PESD, and the full \textsc{Bridge} model. We report Recall and NDCG at cut-offs 5 and 10 to quantify both retrieval coverage and ranking quality.

\begin{table}[h]
  \centering
  \caption{Component-level ablation on \textbf{Beauty}.}
  \label{tab:ablation}
  \setlength{\tabcolsep}{7pt}
  \begin{tabular}{lcccc}
    \toprule
    \textbf{Model}  & \textbf{R@5} & \textbf{N@5} & \textbf{R@10} & \textbf{N@10} \\
    \midrule
    EAR                    & 0.0539 & 0.0396 & 0.0740 & 0.0462  \\
    EAR + CGLR             & 0.0582 & 0.0414 & 0.0818 & 0.0499  \\
    EAR + PESD             & 0.0560 & 0.0406 & 0.0780 & 0.0485  \\
    \textsc{Bridge} (all)        & \textbf{0.0607} & \textbf{0.0443} & \textbf{0.0835} & \textbf{0.0520}  \\
    \bottomrule
  \end{tabular}
\end{table}

\paragraph{Individual contributions}
Adding CGLR to the EAR backbone improves Recall@10 by 10.5\% and NDCG@10 by 8.0\%. This supports our grounding-gap hypothesis: providing $\mathbf{r}$ with a semantic anchor before $s^1$ prediction improves coarse-level decisions without relying on additional supervision. Adding PESD alone to the EAR backbone also yields a consistent gain (Recall@10 +5.4\%, NDCG@10 +5.0\%), suggesting that neighborhood-aware soft targets help even without codebook grounding.

\paragraph{Interaction between CGLR and PESD}
The marginal gain from PESD on Recall@10 is smaller when CGLR is already active (+2.1\% moving from EAR+CGLR to \textsc{Bridge}) than when it is not (+5.4\% from EAR to EAR+PESD), suggesting partial overlap between the two mechanisms on this metric. Overall, combining both components improves EAR by 12.8\% on Recall@10 and 12.6\% on NDCG@10 (EAR $\rightarrow$ \textsc{Bridge}).

\paragraph{Full model}
\textsc{Bridge} achieves the best results across all metrics, indicating that the three gaps are complementary rather than redundant. CGLR improves early coarse decisions by grounding $\mathbf{r}$, PESD improves robustness under self-generated prefixes via neighborhood-aware soft targets, and together they yield consistent gains in both Recall and NDCG beyond either component alone. With EAR's hierarchy-aware compute allocation, this suggests that addressing grounding, neighborhood, and difficulty jointly is necessary to fully exploit the RQ-VAE semantic structure.

\subsection{Analysis of Semantic Quantization Gap}
\label{sec:sqg_analysis}

To assess whether \textsc{Bridge} addresses each gap at the mechanism level rather than merely improving end-to-end recall, we conduct targeted diagnostics for each of the three gaps identified in Section~\ref{sec:intro}.

\paragraph{Grounding gap: early-level SID prediction difficulty}
The grounding gap manifests most directly at the first SID level $s^1$, which carries the highest prediction entropy and determines the coarse semantic region for all subsequent levels. Adding CGLR to EAR+PESD reduces the $s^1$ teacher-forcing cross-entropy loss by 3.95\% on Beauty, suggesting that codebook-grounded latent reasoning helps alleviate difficulty at the hardest, information-scarcest position. Online A/B testing corroborates this finding, with $s^1$ loss consistently reduced by 2--3\% across deployment windows. This asymmetric reduction is consistent with the grounding gap hypothesis: injecting frozen codebook centroids before predicting $s^1$ provides the decoder with the semantic anchoring that the RQ-VAE quantizer encodes but standard SID generation discards.

\begin{figure}[t]
  \centering
  \includegraphics[width=\linewidth]{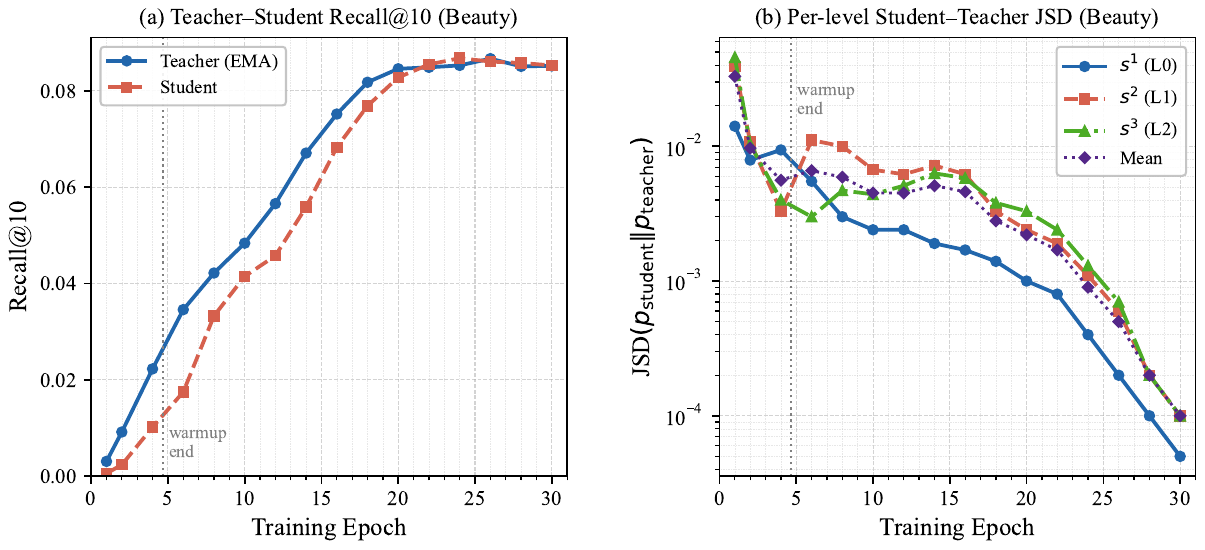}
  \caption{
    Neighborhood gap analysis on Beauty.
    \textbf{(a)}~Teacher and student Recall@10 across training epochs;
     after distillation warm-up, the student closes the gap and reaches
    near-parity by epoch~22.
    \textbf{(b)}~Per-level JSD decays monotonically from 0.0330 to below
    $10^{-4}$, with $s^1$ converging fastest.
  }
  \label{fig:neighborhood_gap}
\end{figure}

\paragraph{Neighborhood gap: knowledge transfer from a privileged teacher}
The neighborhood gap hypothesis predicts that a privileged EMA teacher, which observes semantic neighbor embeddings unavailable to the student, should maintain higher Recall@10 during training, and that PESD should progressively close this gap as distillation ramps up. Figure~\ref{fig:neighborhood_gap}(a) plots teacher and student Recall@10 on the Beauty validation set across training epochs. The teacher leads the student throughout the early and mid-training phases; after the distillation warm-up (step~2{,}000), the student rises steadily and closes the gap, reaching near-parity with the teacher by epoch~22. Beyond this point the student marginally exceeds the teacher on several epochs, which is an expected consequence of EMA weight lag: the student continues to receive gradient updates while the teacher's parameters evolve only through the slow-moving average, so the student's task-specific optimization eventually outpaces the momentum copy. Figure~\ref{fig:neighborhood_gap}(b) quantifies knowledge transfer via the per-level Jensen–Shannon divergence between student and teacher output distributions. Mean JSD begins at 0.0330 before warm-up and decays monotonically to below $10^{-4}$ by epoch~30, with $s^1$ converging fastest, consistent with EAR concentrating recurrent depth at that level. The strictly decreasing JSD trajectory after warm-up suggests that the student progressively absorbs the neighborhood geometry encoded in the teacher's soft targets rather than merely co-evolving with them.

\paragraph{Difficulty gap: compute allocation and inference efficiency}
EAR redistributes compute across SID levels through weight sharing rather than depth scaling: the loop decoder is shared across all levels and all iterations, so trainable parameters are determined solely by the loop block depth $N$ and the single end block, not by the per-level budget $\mathbf{m}$. With $N=3$ loop block layers plus one end block, EAR uses 57.3\% fewer parameters than GR4AD's 9-layer decoder (7 shared + 2 step layers), while the right-aligned schedule $\mathbf{m}=(4,2,1)$ applies the shared layers 4, 2, and 1 times at levels $s^1$, $s^2$, $s^3$ respectively, matching GR4AD's effective depth at the hardest level while substantially reducing compute at easier ones. Online A/B testing against GR4AD on NVIDIA L20 machines shows that EAR improves inference throughput by about 20\% in terms of QPS. These results indicate that EAR helps close the difficulty gap not by adding capacity but by redistributing existing compute according to the quantization-induced level-wise difficulty prior, achieving a favorable accuracy–efficiency trade-off.

\subsection{Online A/B Testing}
We conducted a one-week online A/B test in an e-commerce advertising scenario on a large-scale short-video platform with 5\% of the traffic. The proposed method achieved a 0.65\% eCPM (revenue) uplift, indicating promising business gains and validating the effectiveness of our model.

\section{Conclusion}

We identify the Semantic Quantization Gap (SQG), a mismatch between offline RQ-VAE tokenization and online semantic-ID (SID) generation that weakens semantic grounding, removes neighborhood supervision, and distorts hierarchy-wise compute allocation. We propose \textsc{Bridge}, a shared-loop decoder that combines codebook-grounded latent reasoning, privileged EMA self-distillation, and effort-aware hierarchical recurrence to close SQG. Across three benchmarks, \textsc{Bridge} outperforms strong baselines and ablations indicate the contribution of each component. Future work includes learning the loop schedule end-to-end and extending \textsc{Bridge} to dynamic item sets and multi-modal SIDs.

\section*{Ethical Considerations}
We study generative retrieval for recommendation using public datasets and item-content embeddings. Potential risks include (i) propagating or amplifying biases present in the underlying data and text encoder, (ii) privacy leakage if deployed on sensitive user logs, and (iii) misuse for manipulative ranking. To mitigate these risks in practice, deployments should minimize or anonymize user signals, apply access control and monitoring, and evaluate subgroup performance and calibration to detect disparate impact. Our method does not require additional user-side features beyond standard interaction histories; the privileged neighbor signal used in distillation is computed from item embeddings and is available only during training.

\bibliographystyle{ACM-Reference-Format}
\bibliography{references}

\end{document}